\documentclass[%
 reprint,
 showkeys,
 amsmath,amssymb,
 aps,
]{revtex4-1}

\usepackage{graphicx}% Include figure files
\usepackage{dcolumn}% Align table columns on decimal point
\usepackage{bm}% bold math
\begin{document}

%\preprint{APS/123-QED}

\title{Ignorance can be evolutionarily beneficial}% Force line breaks with \\
%title{Remaining ignorant can have fitness benefits}
\author{Jared M. Field}
  \email{jared.field@maths.ox.ac.uk}
\affiliation{%
Wolfson Centre for Mathematical Biology, Mathematical Institute, University of Oxford, Oxford OX2 6GG,
United Kingdom \\
 }%
\altaffiliation[Also at ]{Mathematical Ecology Research Group, Department of Zoology, University of Oxford, Oxford, UK}
\author{Michael B. Bonsall}
\affiliation{%
Mathematical Ecology Research Group, Department of Zoology, University of Oxford, Oxford, UK \\
 }%
%\affiliation{Mathematical Ecology Research Group, Department of Zoology, University of Oxford, Oxford, UK}
%\author{Supervisor: Mike Bonsall}
%\affiliation{
%\affiliation{
%\affiliation{%


\date{\today}% It is always \today, today,

\begin{abstract}
Information is increasingly being viewed as a resource used by organisms to increase their fitness. Indeed, it has been formally shown that there is a sensible way to assign a reproductive value to information and it is non-negative.  However, all of this work assumed that information collection is cost-free. Here, we account for such a cost and provide conditions for when the reproductive value of information will be negative. In these instances, counter-intuitively, it is in the interest of the organism to remain ignorant. We link our results to empirical studies where Bayesian behaviour appears to break down in complex environments and provide an alternative explanation of lowered arousal thresholds in the evolution of sleep.  
\end{abstract}

%\begin{description}


%\end{description}
%\end{abstract}

%Use showkeys class option if keyword
\maketitle

%\tableofcontents
\section*{}
In all areas of biology observations or cues can elicit a change in behaviour or phenotype. For example, the duration of daylight hours affects the flowering time of plants \cite{amasino2010seasonal}, chemotactic gradients provide a way for bacteria to locate favourable environments \cite{adler1966chemotaxis} and the sighting of a predator may cause an animal to flee. In each case, the observation aids in the choice of an action that will benefit the organism. 
In this way, if there is no cost in collecting information then organisms should always collect it.  Indeed, borrowing from economic theory, but replacing utilities with reproductive values, it has been shown that the reproductive value of information is always non-negative \cite{mcnamara2010information,pike2016general}. This remarkable result suggests that organisms can never decrease their fitness by being more informed. This has far reaching implications in areas of  biology as diverse as public goods games, foraging, collective behaviour and sleep. However, in reality, information comes at a cost \cite{laughlin1998metabolic}. In this paper, we investigate the consequences of formally including such costs in the current theoretical framework. While information may be inherently valuable in decision making, we challenge the view that it always should be or indeed is collected. 

In the context of organismal biology, information use has been approached from two very distinct angles. The first employs the information theory pioneered by Shannon and Weaver \cite{ShannonWeaver}. The alternative, which we take, makes use of statistical decision theory \cite{mcnamara1980application, dall2005information,mcnamara2006bayes}. The former focuses on uncertainty reduction whereas the latter considers how information updating, via Bayes rule, explicitly affects fitnesses. Only recently was the connection between these two approaches shown. Strikingly, mutual information (an information-theoretic measure which quantifies the uncertainty of an outcome after an observation) provides an upper bound on the value of information (a decision-theoretic measure expressed explicitly in terms of fitnesses) \cite{donaldson2010fitness}. However, this is only the case when the fitness measure used is long-term lineage growth rate. Here, we use individual reproductive values. In this way, the value of information is defined by taking the difference in expected optimal reproductive values before and after collecting information. The literature on information use in biology is vast so we do not try to cover it here. However, for useful reviews see \cite{dall2005information,seppanen2007social, valone2007eavesdropping, schmidt2010ecology}

We start by adopting the same framework as in \cite{mcnamara2010information}. Following this, we derive conditions, in the absence of a cost, for when the reproductive value of information is precisely equal to zero. In this case, an organism that collects information will be no better off than one that does not. Next, we account for a cost of collecting information. We do this by discounting the previous reproductive values associated with certain actions and beliefs prior to information collection. This way, the cost is due solely to the collection of information itself. In the proceeding section, we show that the same conditions that we derived previously now lead to the reproductive value of information being negative. In this instance, an organism that remains ignorant will have a higher fitness than one that does not. This has implications for many empirical studies whereby, in increasingly complex environments, organisms were found to stop behaving in a Bayesian manner \cite{j2006animals}. It may be that the cost of collection in these complex environments outweighs any benefits. Additionally, we consider the ramifications for evolutionary problems such as the evolution of sleep. In particular, we suggest that the lowered arousal thresholds associated with sleep, often explained as the by-product of other vital functions, may instead be accounted for by our results \cite{lima2007behavioural}.  We finish by summarising our findings and suggesting future directions of work relating ignorance at the level of the individual to behaviour at the level of the group \cite{couzin2011uninformed}.
%NEED MORE ON BROADNESS
%Mcnarama background. Relation to sleep. What we've done. Ignorance etc. 
\section*{Results}
\subsection*{Framework}
Following \cite{mcnamara2010information}, we start by assuming there are $n$ possible true states of nature, collected in the vector $\Theta = (\theta_1, .., \theta_n)$, about which a given organism is unsure. This organism believes, with probability $p_i$, that $\theta _i$ is the true state. In this way the vector $P = (p_1, ..., p_n)$ summarises the organism's (imperfect) knowledge of nature. 

Further, we assume that this organism must take a certain action, given its beliefs, which will affect its fitness. In particular, a certain action $u$ will have reproductive value $V_i(u)$ given the true state of nature is $\theta_i$. With this set-up, we find the expected reproductive value of action $u$ to be given by 
\begin{equation}
V(u, P) = \sum_{i=1}^n p_iV_i(u). \label{1}
\end{equation}
Then for any vector $P$, we define the optimal action $u^*$ so that 
\begin{equation}
V(u, P) \le V(u^*, P),  \label{2}
\end{equation}
for any action $u$.

We now suppose that this organism can gather information thereby updating its knowledge. We assume it does this by sampling a random variable $X$ which depends on $\Theta$. We denote the probability the observed value of $X$ is $x$ given $\theta_i$ by $f(x|\theta_i)$. With this, we can interpret $p_i$ to be a prior probability and use the observation to form its posterior using Bayes rule, which we denote by $q_i(x)$. Doing so gives
\begin{equation}
q_i(x) = \frac{p_i f(x|\theta_i)}{\sum^n_{j=1}p_jf(x|\theta_j)}.
\label{3}\end{equation}
With this extra information, the organism's knowledge is now summarised by $Q(X) = (q_1, ..., q_n)$. 

With this set-up we define the expected reproductive value (taken over observations $X$) of information $I$ such that \cite{mcnamara2010information}: 
\begin{equation}
I = \mathbf{E}[V(\tilde{u}(X), Q(X))] - V\left(u^*, P\right).
\end{equation}

The first term of this quantity is the reproductive value after collecting information, optimised over actions, and averaged over observations. The second term is the optimal reproductive value if information is not collected. In this way, the difference quantifies the benefit of collecting information.

\subsection*{Expected value of information is non-negative}
While inequality \eqref{2} is expressed in terms of $P$ it is also true for the random vector $Q(X)$. In particular, 
\begin{equation}
V(u, Q(X)) \le  V(\tilde u(Q(X)), Q(X)).
\end{equation}
Taking the expected value with respect to $X$ we then have that
\begin{equation}
\mathbf{E}[V(u, Q(X))] \le \mathbf{E}[ V(\tilde u(Q(X)), Q(X))]. \label{7}
\end{equation}
The left hand side of inequality \eqref{7} can be rewritten as
\begin{align}
\mathbf{E}[ V(u, Q(X))]  &= \mathbf{E}[\sum^n_{i=1}q_i(X) V_i(u)] \label{10.1}\\
&= \sum^n_{i=1}\mathbf{E}[q_i(X)  V_i(u)] \label{10.2} \\
&=  \sum^n_{i=1}\mathbf{E}[q_i(X) ] V_i(u) \label{10}
\end{align}
However, as the expectation is taken over observations $x$, we have
\begin{equation}
\mathbf{E} [q_i(X)] = \sum _x q_i(x) f(x) \label{11}
\end{equation}
where $f(x)$ is the probability that the observed value of $X$ is $x$.  Using the law of total probability, $f(x)$ can in fact be written as 
\begin{equation}
f(x) = \sum^n_{j=1}p_jf(x|\theta_j), \label{tot}
\end{equation}
so that,  coupled with \eqref{3}, \eqref{11} can be reexpressed as
\begin{align}
\mathbf{E} [q_i(X)] &= p_i \sum _x  f(x|\theta_j), \\
&= p_i, \label{14}
\end{align}
as $f(x|\theta_j)$ is a probability distribution. 

Now, using \eqref{14} in \eqref{10} we find that 
\begin{align}
\mathbf{E}[V(u, Q(X))] &= \sum_{i=1}^n p_iV_i(u) \\
&= V(u, P), \label{15}
\end{align}
by \eqref{1}. This statement is true for any action $u$ so that it is in particular true of the action $u^*(P)$ that optimises $V(u, P)$. 

Finally, using \eqref{15} in inequality \eqref{7} and rearranging we find that 
\begin{equation}
0 \le \mathbf{E}[ V(\tilde u(Q(X)), Q(X))] - V(u^*, P).
\end{equation}
Otherwise put, the expected reproductive value of information $I$ is non-negative.

\subsection*{When is the expected value of information equal to zero?}
So far we have shown (and also in \cite{mcnamara2010information}) that the expected value of information is non-negative. However, if it is equal to zero then an organism that collects information will be no better off than an organism that does. 

To investigate this, we start by bounding inequality \eqref{7} from above. Similarly to equations \eqref{10.1} and  \eqref{10.2} but for the right hand side of \eqref{7} we have 
\begin{align}
\mathbf{E}[ V(\tilde u(Q(X)), Q(X))] &= \mathbf{E}[\sum^n_{i=1}q_i(X) V_i(\tilde u(X)] \\
&= \sum^n_{i=1}\mathbf{E}[q_i(X)  V_i(\tilde u(X))].
\end{align}
However this time as the action $u(X)$ taken depends on the observation we must take the expectation directly. In particular, we have 
\begin{equation}
\mathbf{E}[ V(\tilde u(Q(X)), Q(X))] = \sum^n_{i=1} \sum_x q_i(x)  V_i(\tilde u(x)) f(x).
\end{equation}
We now suppose that there exists some $\hat x$ such that 
\begin{align}
\sum^n_{i=1} \sum_x q_i(x)  V_i(\tilde u(x)) f(x) \le \sum^n_{i=1} \sum_x q_i(x)  V_i(\tilde u(\hat{x})) f(x) \label{20}
\end{align}
Using \eqref{3}, this then becomes
\begin{align}
\sum^n_{i=1} \sum_x q_i(x)  V_i(\tilde u(x)) f(x) &\le  \sum^n_{i=1} p_i V_i(\tilde u(\hat{x})) \sum_x f(x|\theta_i) \\
&= V(\tilde{u}(\hat{x}), P), 
\end{align}
as $f(x|\theta_i)$ is a probability distribution. Hence, if such an $\hat{x}$ does exist, we have showed that 
\begin{equation}
V(u^*, P) \le \mathbf{E}[ V(\tilde u(Q(X)), Q(X))] \le V(\tilde{u}(\hat{x}), P).
\end{equation}
However, from the very definition of $u^*$ it must be that
\begin{eqnarray}
V(\tilde{u}(\hat{x}), P) \le V(u^*, P),
\end{eqnarray}
so that in fact 
\begin{eqnarray}
\mathbf{E}[ V(\tilde u(Q(X)), Q(X))] = V(u^*, P),
\end{eqnarray}
and so 
\begin{equation}
I = 0.
\end{equation}

For such an $\hat{x}$ to exist we need, from \eqref{20}, that 
\begin{equation}
 \sum_x q_i(x)  V_i(\tilde u(x)) f(x) \le  \sum_x q_i(x)  V_i(\tilde u(\hat{x})) f(x),
\end{equation}
for each $i$. Using \eqref{3} this is equivalent to requiring 
\begin{align}
\sum_x f(x|\theta_i)  V_i(\tilde u(x))  &\le V_i(\tilde u(\hat{x})) \sum_x f(x|\theta_i) \\
&= V_i(\tilde u(\hat{x})),
\end{align}
which is in turn equivalent to 
\begin{equation}
f(\hat x|\theta_i)  V_i(\tilde u(\hat x)) + \sum_{x\ne \hat{x}}f(x|\theta_i)  V_i(\tilde u(x)) \le V_i(\tilde u(\hat{x})).
\end{equation}
This statement will clearly be true in three cases. Either 
\begin{equation}
 V_i(\tilde u(x)) = 0, \label{31}
\end{equation}
for each $x \ne \hat{x}$ and for each $i$ or
\begin{equation}
f(x|\theta_i) = 0, \label{32}
\end{equation}
for each $x \ne \hat{x}$ and for each $i$ or, finally, a mixture of these two previous extreme cases. In particular, we can have instances where there is a particular $x\ne \hat{x}$ so that 
\begin{equation}
V_i(\tilde u(x)) \ne 0, f(x|\theta_i) = 0, \label{33}
\end{equation} 
for each $i$ or \emph{vice versa} so that
\begin{equation}
 V_i(\tilde u(x)) = 0, f(x|\theta_i) \ne 0 , \label{34}
\end{equation}
for each $i$.

Condition \eqref{31} asserts that all other possible actions will have a reproductive value of zero. Though unlikely, this may be realised if presented with an extreme situation (such as certain death in the face of predation) when only one action will lead to survival. 

Alternatively, via \eqref{tot} condition \eqref{32} implies 
\begin{equation}
f(\hat{x}) = 1.
\end{equation}
The interpretation here is that only the observation $\hat{x}$ will be observed. In this way, this condition says something about the environment in which an organism finds itself. If the organism finds itself in a period of relative constancy (such as certain safety due to a temporary lack of predators), this condition will be fulfilled. 

The blended conditions \eqref{33} and \eqref{34} represent the more realistic situations where there is still flexibility in actions with nonzero reproductive value but the observation to which it corresponds will not be observed or where an observation will have nonzero probability of being observed but the associated reproductive value is zero, respectively. Otherwise put, an organism may be adapted and able to respond to a certain cue however currently is in a situation where that cue will not be observed. Alternatively, the organism may observe a certain cue but is unable to adequately respond.  
\subsection*{Accounting for a cost of information collection}
We now assume that the collection of information comes at a cost $c$. The reproductive value associated with any given action $u$ taken after gathering information will then be discounted by this cost. Hence, if an organism collects information the reproductive value of action $u$ given state $\theta_i$ will be
\begin{equation}
\bar{V}_i(u) = V_i(u) - c. \label{4}
\end{equation}
As before, we have that 
\begin{equation}
\mathbf{E}[\bar V(u, Q(X))] \le \mathbf{E}[ \bar V(\tilde u(Q(X)), Q(X))], 
\end{equation}
for any $u$. Replacing $V_i(u)$ with $\bar V_i(u)$ in \eqref{10}, the left hand side of this inequality can be written as
\begin{equation}
\mathbf{E}[\bar V(u, Q(X))]=V(u, P) - c, 
\end{equation}
which is, in particular, true for $u^*$ so that we have 
\begin{equation}
V(u^*, P) - c \le \mathbf{E}[ \bar V(\tilde u(Q(X)), Q(X))]
\end{equation}
Similarly, replacing occurrences of $V_i(\cdot)$ with $\bar V_i(\cdot)$ in the entire preceding section it follows that 
\begin{equation}
\mathbf{E}[ \bar V(\tilde u(Q(X)), Q(X))] \le V(u^*, P)-c,
\end{equation} 
so that, if there is a cost to collecting information we find that 
\begin{equation}
\mathbf{E}[ \bar V(\tilde u(Q(X)), Q(X))] = V(u^*, P)-c.
\end{equation}
Hence
\begin{equation}
I = -c.
\end{equation}
This result, as before, relies on one of the four conditions \eqref{31}-\eqref{34} being true. In this case, an organism that collects information will be worse off than an organism that does not.  

\subsection*{Alternative measures for valuing information}
With some important recent exceptions, it is not generally believed that organisms are behaving in a strictly Bayesian manner \cite{j2006animals, biernaskie2009bumblebees,louapre2011information}. Instead, it is suspected that they follow Rules of Thumb that approximate optimal Bayesian strategies \cite{mcnamara1980application,mcnamara2006bayes}. These rules are based on either the experience of an organism's ancestors (and so is genetically encoded), the experience of the organism itself or a combination of both \cite{giraldeau1997ecology,j2006animals, mcnamara2006bayes} . When these rules are based on the experience of the organism they will be informed by typical observations. In some circumstances, the expected value (such as with the analysis performed above) will be a good indication of typical sampled values. 

However, in other cases the expected value is in fact a very poor measure of typical values. To be more concrete, consider $G(X)$ defined such that 
\begin{equation}
 G(X) = V(\tilde u(Q(X)), Q(X)) - V(u^*, P)
\end{equation}
with probability density function $g(x)$. Note that this random variable has the property that $\mathbf{E}[G(X)] = I$. 

If $g(x)$ is a unimodal symmetric distribution then $I$ will be a decent indication of typical values. However, if $g(x)$ is, say, multimodal or skewed then $I$ will be a poor measure. Moreover, if $g(x)$ is positively skewed then typical values of $G$ may be negative. In this case, \emph{even without a cost associated with the act of information collection}, it will not be beneficial for an organism operating under a Rule of Thumb to collect information. 

Whether or not $g(x)$ is skewed will depend on $f(x)$, the distribution describing $X$. However, the random variable $X$ describes the environment the organism is in. For this reason, it is highly plausible that $X$ and so $g(x)$ will change as a function of time. For example, it may be that predators are more prevalent at a certain time over the course of one day. In this case, the probability of observing a predator will also change over that period. So too, then, will $g(x)$.

\section*{Discussion}
The notion of putting a value on information has existed in economic theory for quite some time \cite{gould1974risk}. Despite this, analogous work in the context of animal behaviour has taken a little longer to catch up. Recently, however, it was formally shown that the reproductive value of information is always non-negative when there is no cost in its collection \cite{mcnamara2010information}. From this, it has been concluded that organisms should always collect information. This conclusion has since not been investigated much further in the literature. While costs associated with sampling have been considered in some evolutionary games (see \cite{mcnamara2009evolution}, where  the costs are crucial to maintain a mix of strategies), the focus has not been on the whether or not information should be collected in the first place. %In particular, any costs associated with the collection of information have been largely ignored. 

Here, using the same framework, we derived particular conditions for when the value of information will be equal to zero. In such a case, an organism that collects information will in fact be no better off than one that does not. Following this, we explicitly accounted for a cost of collecting information. In reality, this may come in the form of energy, time or resources that could otherwise be spent on other vital biological functions \cite{laughlin1998metabolic}. In this instance, we found that under the same conditions as before, the value of information will now be negative. Thus, there will be times when the collection of information will have a negative impact on an organism's fitness. 

These conditions can be broadly organised into three groups. First, if an organism finds itself (potentially temporarily) in a situation whereby only one action will lead to a non-zero reproductive value, then the value of information will be negative. Second, if it is such that only one observation will be observed (again, potentially temporarily), then an organism will not benefit from collecting information. Both of these cases, in the absence of a cost of collecting information, are emphasised informally in \cite{pike2016general}. The third, a mixture of the previous two conditions, is not. In this case, an organism can still be flexible in its actions, and there can still be variance in observations, and yet the value of information will be less than zero. To be more concrete, this third condition will be fulfilled if an organism may observe a certain cue but is unable to adequately respond or if it is adapted and able to respond to a certain cue however currently is in a situation where it will not be observed. In either case, an organism will do better if they remain relatively ignorant. 

Though the results presented here are quite broad they, and in particular the last condition, have strong implications for the evolution of sleep and sleep-like states. Broadly speaking, sleep can be defined physiologically (characterised by certain brain activities) or behaviourally (characterised by inactivity and arousal thresholds). Some organisms, like dolphins,  sleep according to one definition but not the other \cite{siegel2009}. It is however the behavioural definition, and in particular the lowered arousal thresholds, that presents the evolutionary puzzle. This vulnerable disconnect from an organism's environment is most often explained by assuming \emph{a priori} that sleep serves some vital function for which this behavioural shutdown is necessary \cite{lima2007behavioural}. Here, however, we have shown that such an assumption is not necessary. If there is a cost to collecting information, then there will be times when it should not be collected. We are not, of course, suggesting that vital functions of sleep do not exist. However, our result opens up the possibility that the vital functions evolved after behavioural shutdowns and not the other way around.   

In an empirical review of animal Bayesian updating, it was found that in simple environments all but one species performed consistently with Bayesian predictions \cite{j2006animals}. In complex environments, this was not found to be true. The explanation put forward was that in these complex environments it is more difficult, or takes longer, for the organism in question to successfully learn prior distributions. An alternative and complementary explanation, suggested by our results, is that the cost of collecting information may be too high in these environments. In other words, they are making a Bayesian decision to remain ignorant. It would be exceedingly interesting to test this hypothesis experimentally. 

More generally, we have shown that periods of ignorance can lead to fitness benefits at the level of the individual. For future theoretical work, it would be fruitful to understand how this may translate to behaviour at the level of the group. Recently, it has been shown how uninformed individuals can help democratic consensus be reached in the face of internal conflicts \cite{couzin2011uninformed}. This particular study however makes no reference to individual fitnesses but is, in essence, mechanistic. It would be particularly interesting to see if, taking an evolutionary approach, similar conclusions can be found.  

\section*{Competing interests}
We have no competing interests.
\section*{Authors' contributions}
JMF carried out the research. JMF and MBB wrote the manuscript.
\section*{Funding}
JMF is funded by the Charles Perkins Scholarship with additional financial support from UTS, Sydney.
\section*{Acknowledgments} 
We thank Thomas W. Scott and John M. McNamara for valuable comments and discussion.
\bibliographystyle{plain}
\bibliography{ref_2017}
\end{document}